\documentclass[11pt]{article}
\usepackage{amsmath}
\usepackage{CJK}
\usepackage{subfigure}
\usepackage{graphicx}
\usepackage{dsfont}\usepackage{color}
\usepackage{amscd}
\newtheorem{thm}{Theorem}

 \newtheorem{prop}{Proposition}

\begin{document}
\title{Nonlinear Schr\"odinger Equation: Generalized Darboux Transformation and Rogue Wave Solutions}

\date{}
\author{Boling Guo$^\dag$, Liming Ling$^\dag$ and Q. P. Liu$^\ddag$\\
$^\dag$Institute of Applied Physics and Computational Mathematics\\
Beijing 100088, P R China\\[10pt]
and\\
$^\ddag$Department of Mathematics\\
China University  of Mining and Technology\\
Beijing 100083, P R China}
\maketitle


\begin{abstract}
In this paper, we construct a generalized Darboux transformation for
nonlinear Schr\"odinger equation. The associated $N$-fold Darboux
transformation is given both in terms of a summation formula and in
terms of determinants. As applications, we obtain compact
representations for the $N$-th order rogue wave solutions of the
focusing  nonlinear Schr\"odinger equation and Hirota equation. In particular, the dynamics of the general third order rogue wave is discussed and  shown to exhibit interesting structure.
\end{abstract}

\maketitle
\section{Introduction}
Darboux transformation, originated from the work of Darboux in 1882 for the Sturm-Liouville equation,
is a powerful method to construct solutions for integrable systems. The theory is presented by several monographs and review papers (see \cite{C,Mat1,DL}).  In literature, various approaches have been proposed to find a Darboux transformation for a given equation, for instance, the operator factorization method \cite{AM}, the
gauge transformation method \cite{Neu-M,LGZ,DL}, the loop group
transformation \cite{TU} and so on.

It is remarked that Darboux transformation is very efficient to construct soliton solutions. Indeed, through iterations, one is often leaded to compact representations in terms of special determinants such as Wronskian or Grammian for $N-$soliton solutions. Such $N-$soliton solutions are appealing both form the theoretical viewpoint and  form practical application viewpoint as well.

In addition to the soliton solutions, rational solutions are also interesting and Darboux transformation may be adopted for this purpose. For the celebrated Koteweg-de Vries (KdV) equation, Matveev \cite{matveev92} introduces the so-called generalized Darboux transformation and the positon solutions are calculated.  Recently, rogue waves, appeared in oceans, have been studied and applied extensively in other fields  such as
Bose-Einstein condensates, optics and superfluids and so on. The very first model for rogue waves is the focusing nonlinear Schr\"{o}dinger (NLS) equation
\begin{equation}\label{nls}
  {\rm i}q_t+\frac{1}{2}q_{xx}+|q|^2q=0,
\end{equation}
which  has been an important integrable equation. While the simplest rogue wave solutions are calculated by Akhemidiev and coworkers, and the construction of higher order analogue is one of the challenges as remarked in \cite{ACA}. In this regard, as pointed out by Dubard {\em{et al}} \cite{DPKM}, the solutions, obtained by Eleonskii, Krichever and Kulagin \cite{ekk}, represent a class of multi-rogue wave solutions. It is remarked that the construction method proposed in \cite{DPKM,DM} is very specific and technical, so may not be easy to apply to other models.

The aim of the paper is to propose a simple method to construct multi-rogue wave solutions. The main tool is the generalized Darboux transformation.  We will reexam the Matveev's generalized Darboux transformation for the KdV equation and derive it in a way that could be easily extended to other models. Then, we apply the idea to the focusing NLS equation and work out a formula to generate multi-rogue wave solutions.

This paper is organized as following: In section two, we propose a new way to derive  the generalized Darboux transformation for KdV equation.  In the section three, we first apply the proposed method to the NLS equation case and obtain the corresponding generalized Darboux transformation for it.
 Then, we reformulate the $N$-th fold generalized Darboux transformation in terms of determinants.
  Also, we provide formulae for the $ N$-th order rogue wave solutions for NLS equation and Hirota equation. With the help of this formulae, we consider the dynamics of the general third order rogue wave. The spatial-temporal pattern of the solution could form as a triangle or pentagon. Final section contains some discussions.

\section{Generalized Darboux transformation for KdV}

Let us first recall the well known classical Darboux transformation for the KdV equation. Consider the Sturm-Liouville equation
\begin{equation}\label{S-L}
    -\Psi_{xx}+u\Psi=\lambda\Psi,
\end{equation}
and introduce the following first order operator
\begin{equation*}\label{cdt}
    T[1]=\partial_x-\frac{\Psi_{1x}}{\Psi_1},
\end{equation*}
where $\Psi_1$ is the fixed solution of \eqref{S-L} with
$\lambda=\lambda_1$. Then Darboux transformation
\begin{equation*}
\Psi[1]=T[1]\Psi=\frac{\mathrm{Wr}(\Psi_1,\Psi)}{\Psi_1},
\end{equation*}
converts the equation \eqref{S-L} into
\begin{equation}\label{S-L1}
-\Psi[1]_{xx}+u[1]\Psi[1]=\lambda\Psi[1]
\end{equation}
where
\begin{equation*}
u[1]=u-2(\ln{\Psi_1})_{xx}.
\end{equation*}
and $\mathrm{Wr}(\Psi_1,\Psi)=\Psi_1\Psi_x-\Psi_{1,x}\Psi$ is the
standard Wronskian determinant.

 The most interesting point here
is that one could iterate the above Darboux transformation. Indeed,
the $N$-times iterated or $N$-fold Darboux transformation yields Crum theorem
\begin{equation*}\label{curm}
    -\Psi_{xx}[N]+u[N]\Psi[N]=\lambda \Psi[N], \quad
    u[N]=u-2(\ln{\mathrm{Wr}(\Psi_1,\cdots,\Psi_N)})_{xx},
\end{equation*}
where
$$\Psi[N]=\frac{\mathrm{Wr}(\Psi_1,\cdots,\Psi_N,\Psi)}{\mathrm{Wr}(\Psi_1,\cdots,\Psi_N)},$$
and $\Psi_1,\cdots,\Psi_N$ are solutions of \eqref{S-L} at
$\lambda=\lambda_1,\cdots,\lambda_N$, respectively.

It is obvious that $\Psi_1[1]=T[1]\Psi_1=0$, namely $\Psi_1$ is mapped to   trivial solution. This fact implies that  a seed solution may not be used more than once when considering the iterations for Darboux transformation. However, as pointed out by Matveev and Salle\cite{Mat1}, a generalized Darboux transformation does exist. Let us derive this result in a  way which may be readily generalized. To precede, we assume
that
\[
\Psi_2=\Psi_1(k_1+\epsilon),
\]
where $k_1=f(\lambda_1)$ is a monotonic function and $\epsilon$ is a small parameter.
Expanding $\Psi_2$ in a series in $\epsilon$
\begin{equation*}\label{series}
    \Psi_2=\Psi_1+\Psi_1^{[1]}\epsilon+\Psi_1^{[2]}\epsilon^2+\cdots,
\end{equation*}
where  $\Psi_1^{[i]}=\frac{1}{i!}\frac{\partial^i \Psi_1(k)}{\partial
k^i}|_{k=k_1}$.
Since $\Psi_2[1]\equiv T_1[1]\Psi_2$ is a special solution for \eqref{S-L1}, so is
$\frac{\Psi_2[1]}{\epsilon}$. Taking limit $\epsilon\rightarrow 0$
for this solution, we find
\begin{equation*}\label{limit}
    \Psi_1[1]=\lim_{\epsilon\rightarrow
    0}\frac{T[1]\Psi_1(k_1+\epsilon)}{\epsilon}=T[1]\Psi_{1}^{[1]},
\end{equation*}
which is a non-trivial solution for \eqref{S-L1} at
$\lambda=\lambda_1$. This solution may be adopted to do the second
step Darboux transformation, that is
\begin{equation*}\label{dt2}
    T[2]=\partial_x-\frac{\Psi_{1,x}[1]}{\Psi_1[1]},\quad u[2]=u[1]-2(\ln{\mathrm{Wr}(\Psi_1,\Psi_{1}^{[1]})})_{xx}.
\end{equation*}
Combining these two Darboux transformations, we obtain
\begin{equation*}\label{twostep}
    -\Psi_{xx}[2]+u[2]\Psi[2]=\lambda\Psi[2],\quad
    \Psi[2]=\frac{\mathrm{Wr}(\Psi_1,\Psi_{1}^{[1]},\Psi)}{\mathrm{Wr}(\Psi_1,\Psi_{1}^{[1]})}.
\end{equation*}

This process may be continued and results in the so called
generalized Darboux transformation for the system \eqref{S-L}.
Indeed, let $\Psi_1$, $\Psi_2$,
$\cdots$, $\Psi_n$ be $n$ different solutions for \eqref{S-L} at $\lambda_1, \lambda_2,
\cdots, \lambda_n$, and consider the expansions
\begin{eqnarray*}
  \Psi_i(k_i+\epsilon) &=& \Psi_i(k_i)+\Psi_i^{[1]}\epsilon+\cdots+\Psi_i^{[m_i]}\epsilon^{m_i}+\cdots,\quad k_i=f(\lambda_i)
  \quad (i=1,2,\cdots,n)
\end{eqnarray*}
then we have the following
\begin{prop}{\text{\cite{Mat-g}}}
\begin{equation*}
u[N]=u-2(\ln(W_1))_{xx},\quad\Psi[N]=\frac{W_2}{W_1}
\end{equation*}
with
\begin{align*}
 &W_1=\mathrm{Wr}(\Psi_1,\cdots,\Psi_{1}^{[m_1]},\Psi_2,\cdots,\Psi_2^{[m_2]},\cdots,\Psi_n,\cdots,\Psi_2^{[m_n]}),\\
 &W_2=\mathrm{Wr}(\Psi_1,\cdots,\Psi_{1}^{[m_1]},\Psi_2,\cdots,\Psi_2^{[m_2]},\cdots,\Psi_n,\cdots,\Psi_2^{[m_n]},\Psi),
\end{align*}
solve
\begin{equation*}\label{mcdt}
    -\Psi_{xx}[N]+u[N]\Psi[N]=\lambda \Psi[n],\quad
\end{equation*}
where $m_1+m_2+\cdots+m_n=N-n$, $m_i\geq 0,$ $m_i\in\mathds{Z}$.
\end{prop}

The generalized Darboux transformation presented above may be used to generate both solitions and rational solutions for the KdV equation. Let us illustrate this with the following examples. It is
well known that, the KdV equation
\begin{equation}\label{kdv}
    u_t-6uu_x+u_{xxx}=0,
\end{equation}
takes \eqref{S-L} as its  spatial part of the spectral problem and the corresponding temporal part reads as
\begin{equation*}\label{tpart}
    \Psi_t=-4\Psi_{xxx}+6u\Psi_x+3u_x\Psi.
\end{equation*}

In the case of $N$ distinct spectral parameters, we will have the Wronskian representation for the  $N-$soliton solution. To get rational solutions, one starts with the the seed solution
$u=c$, $c$ is a real constant, and
$\Psi_1=\sin[k_1(x+(4k_1^2+6c)t)+P(k_1)]$, $k_1=\sqrt{\lambda_1-c}$,
$P(k_1)$ is a polynomial of $k_1$. Now expanding the function $\Psi_1$ at
$k_1=0$ and taking $P(k_1)=0$ for convenience, we  have
\begin{equation*}\label{series1}
    \Psi_1=(x+6ct)k_1+\left[-\frac{1}{6}(x+6ct)^3+4t\right]k_1^3+\left[\frac{1}{120}(x+6ct)^5-2t(x+6ct)^2\right]k_1^5+\cdots,
\end{equation*}
therefore
\begin{displaymath}
\Psi_1^{[0]}
=x+6ct,\quad \Psi_1^{[1]}=-\frac{1}{6}(x+6ct)^3+4t,\quad\Psi_1^{[2]}=\frac{1}{120}(x+6ct)^5-2t(x+6ct)^2
\end{displaymath}
then the generalized Darboux transformation provides us the following rational solution for the KdV equation \eqref{kdv}
\begin{equation*}\label{threeration}
    u[3]=c+\frac{G}{H^2},
\end{equation*}
where
\begin{eqnarray*}
  G &=&  12[279936c^5t^5(216c^5t^5+360c^4xt^4+270c^3x^2t^3+120c^2x^3t^2+35cx^4t+7x^5)\\
    &&+38880c^4t^4(180t^2+7x^6)+25920c^3xt^3(x^6+180t^2)+1620c^2x^2t^2(x^6+720t^2)\\
    &&+60ct(x^9+2160t^2x^3+4320t^3)+43200t^3x+x^{10}+5400x^4t^2],\\
  H &=&3888c^4t^4(12c^2t^2+12cxt+5x^2)+4320c^3t^3(x^3+3t)\\
    &&+540c^2xt^2(x^3+12t)+36cx^2t(x^3+30t)+x^6-720t^2+60x^3t.
\end{eqnarray*}
More general solutions of rational type may be obtained if we expand
\begin{equation*}\label{serieskdv}
    \Psi_1=\sin[k_1(x+(4k_1^2+6c)t)+P(k_1)]=\Psi_1^{[0]}k_1+\Psi_1^{[1]}k_1^3+\Psi_1^{[2]}k_1^5+\cdots+\Psi_1^{[N]}k_1^{2N+1}+\cdots.
\end{equation*}
In particular, the positon solutions for the KdV equation may be found this way and for a detailed analysis we refer to the work of Matveev \cite{Matveev}.

Concluding this section, we mention that, for those equations which possess solutions represented in terms of Wronskians, there are alternative manners to explain the limit solutions \cite{Mat1,MLH,CWLZ}.

\section{Generalized Darboux transformation for NLS}
In this section, we extend the idea discussed above section to the NLS equation and construct a  generalized Darboux transformation for it. Furthermore, we shall show that such Darboux transformation enables one to obtain, apart from the soliton solutions, rational solutions including multi-rogue wave
 solutions.

The focusing NLS equation, equation \eqref{nls}
is the compatibilty condition of the following linear spectral problems
\begin{subequations}\label{nlsl}
\begin{eqnarray}
\label{nls_x}  \Psi_x &=& [{\rm i}\zeta \sigma_1+{\rm i} Q]\Psi, \\
\label{nls_t}  \Psi_t &=& \left[{\rm i}\zeta^2 \sigma_1+{\rm i}\zeta Q+\frac{1}{2}\sigma_1(Q_x-{\rm
  i}Q^2)\right]\Psi,
\end{eqnarray}
\end{subequations}
where
\begin{equation*}
    \sigma_1=\begin{pmatrix}
               1 & 0 \\
               0 & -1 \\
             \end{pmatrix},\quad Q=\begin{pmatrix}
                                     0 & q^* \\
                                     q & 0 \\
                                   \end{pmatrix}.
\end{equation*}
\subsection{Generalized Darboux transformations}

The Darboux transformation in this case is defined as (see \cite{C} and the references there)
\[
\Psi[1]=T[1] \Psi, \quad q[1]=q+2(\zeta_1^*-\zeta_1)(P[1])_{21},
\]
where
\begin{equation}\label{nlsdt}
    T[1]=\zeta-\zeta_1^*+(\zeta_1^*-\zeta_1)P_1[1],\quad P[1]=\frac{\Psi_1\Psi_1^{\dag}}{\Psi_1^{\dag}\Psi_1}
\end{equation}
and  $\Psi_1$ is a special solution of the linear system \eqref{nls_x}-\eqref{nls_t} at
$\zeta=\zeta_1$, $(P_1[1])_{21}$ represents the entry of matrix $P_1[1]$ of the second row and first column and $``\dag"$ denotes the matrix transpose and
complex conjugation.

If $N$ distinct seed solutions $\Psi_k, (k=1,2,...,N)$ of
\eqref{nls_x}-\eqref{nls_t} are given, the basic Darboux
transformation may be iterated. To do the second step of transformation, we employ $\Psi_2$ which is mapped to $\Psi_2[1]=T[1]|_{\zeta=\zeta_2}\Psi_2$. Therefore,
\[
\Psi[2]=T[2]\Psi[1], \quad q[2]=q[1]+2(\zeta_2^*-\zeta_2)(P[2])_{21},
\]
where
\[
T[2]=\zeta-\zeta_2^*+(\zeta_2^*-\zeta_2)P[2],\quad
P[2]=\frac{\Psi_2[1]\Psi_2[1]^{\dag}}{\Psi_2[1]^{\dag}\Psi_2[2]}.
\]
In general case, we may have
\begin{thm}
Let $\Psi_1$, $\Psi_2$, $\cdots$,
$\Psi_N$ be $N$ distinct solutions of the spectral problem \eqref{nls_x}-\eqref{nls_t} at
$\zeta_1, \cdots, \zeta_N$ respectively, then the N-fold Darboux transformation for NLS equation \eqref{nls} is
\begin{equation}\label{n-fold}
\Psi[N]= T[N]T[N-1]\cdots T[1]\Psi,\quad q[N]=q[0]+2\sum_{i=1}^{N}(\zeta_i^*-\zeta_i)(P[i])_{21}
\end{equation}
with
\begin{eqnarray*}
  &&T[i] = \zeta-\zeta_i^*+(\zeta_i^*-\zeta_i)P[i],\quad P[i]=\frac{\Psi_i[i-1]\Psi_i[i-1]^{\dag}}{\Psi_i[i-1]^{\dag}\Psi_i[i-1]}, \\
  &&\Psi_i[i-1] = \left(T[i-1]T[i-2]\cdots T[1]\right)|_{\zeta=\zeta_i}\Psi_i,\quad q[0]=q.
\end{eqnarray*}
\end{thm}
We remark that the $N$-fold Darboux transformation given by \eqref{n-fold} is equivalent to the determinant representation presented in \cite{Mat1} as will be seen in the next subsection.

Now we manage to find a generalized
Darboux transformation. As in last section, suppose that
$\Psi_2=\Psi_1(\zeta_1+\delta)$ is a special
 solution for system, then after transformation we have $\Psi_2[1]=T_1[1]\Psi_2$. Expanding $\Psi_2$ at $\zeta_1$, we have
\begin{equation}\label{expand}
    \Psi_1(\zeta_1+\delta)=\Psi_1+\Psi_1^{[1]}\delta+\Psi_1^{[2]}\delta^2+\cdots+\Psi_1^{[N]}\delta^N+\cdots,
\end{equation}
where $\Psi_1^{[k]}=\frac{1}{k!}\frac{\partial^k}{\partial \zeta^k}\Psi_1(\zeta)|_{\zeta=\zeta_1}$.

Through the following limit process
\begin{equation}\label{nlsn1}
\lim_{\delta\rightarrow0}\frac{\left[T_1[1]|_{\zeta=\zeta_1+\delta}\right]\Psi_2}{\delta}=\lim_{\delta\rightarrow0}\frac{[\delta+T_1[1]|_{\zeta=\zeta_1}]\Psi_2}{\delta}
=\Psi_1+T_1[1]|_{\zeta=\zeta_1}\Psi_{1}^{[1]}\equiv \Psi_1[1]
\end{equation}
we find a solution to the linear system \eqref{nls_x}-\eqref{nls_t}
with $q[1]$ and $\zeta=\zeta_1$. This allows us to go to the next step
of Darboux transformation, namely
\begin{equation}\label{nlsdt2}
    T_1[2]=\zeta-\zeta_1^*+(\zeta_1^*-\zeta_1)P_1[2],\quad
    q[2]=q[1]+2(\zeta_1^*-\zeta_1)(P_1[2])_{21},
\end{equation}
where
\[
P_1[2]=\frac{\Psi_1[1]\Psi_1[1]^{\dag}}{\Psi_1[1]^{\dag}\Psi_1[1]}.
\]

Similarly, the limit
\begin{eqnarray*}
\lefteqn{   \lim_{\delta\rightarrow0}\frac{[\delta+T_1[2](\zeta_1)][\delta+T_1[1](\zeta_1)]\Psi_2}{\delta^2}
    =}\\
    &&\Psi_1+\left[T_1[1](\zeta_1)+T_1[2](\zeta_1)\right]\Psi_{1}^{[1]}+T_1[2](\zeta_1)T_1[1](\zeta_1)\Psi_{1}^{[2]} \equiv \Psi_1[2],
\end{eqnarray*}
provides us a
  nontrivial solution for the linear spectral problem with $q=q[2]$ and $\zeta=\zeta_1$. Thus we may do the third
  step iteration of Darboux transformation, which is
following
\begin{equation}\label{nlsdt3}
    T_1[3]=\zeta-\zeta_1^*+(\zeta_1^*-\zeta_1)P_1[3],\quad
    P_1[3]=\frac{\Psi_1[2]\Psi_1[2]^{\dag}}{\Psi_1[2]^{\dag}\Psi_1[2]},\quad q[3]=q[2]+2(\zeta_1^*-\zeta_1)(P_1[3])_{21}.
\end{equation}

Continuing above process and combining  all the Darboux
transformation, a generalized Darboux transformation is constructed.
We summarize our finding as
\begin{thm}
Let  $\Psi_1(\zeta_1)$, $\Psi_2(\zeta_2)$, $\cdots$,
$\Psi_n(\zeta_n)$ be $n$ distinct solutions of  the linear spectral
problem \eqref{nls_x}-\eqref{nls_t} and
\[
  \Psi_i(\zeta_i+\delta)=\Psi_i+\Psi_i^{[1]}\delta+\Psi_i^{[2]}\delta^2+\cdots+\Psi_i^{[m_i]}\delta^N+\cdots,\quad
  (k=1,2,\cdots,n)
\]
be their expansions, where
 \[
 \Psi_i^{[j]}=\frac{1}{j!}\frac{\partial^j}{\partial
\zeta^j}\Psi_i(\zeta)|_{\zeta=\zeta_i},\quad
  (i=1,2,\cdots,n,j=1,2,\cdots).
 \]
 Defining
 \begin{equation}\label{nlsdtn}
    T=\Gamma_n\Gamma_{n-1}\cdots\Gamma_1\Gamma_0,\quad \Gamma_i=T_i[m_i]\cdots
    T_i[1],\;(i\geq1),\quad \Gamma_0=I,
\end{equation}
where
\begin{eqnarray*}
       T_i[j] &=& \zeta-\zeta_i^*+(\zeta_i^*-\zeta_i)P_i[j],\quad P_i[j]=\frac{\Psi_i[j-1]\Psi_i[j-1]^{\dag}}{\Psi_i[j-1]^{\dag}\Psi_i[j-1]} ,\quad 1\leq j\leq m_i\\
       \Psi_i[0]&=&\left(\Gamma_{i-1}\cdots\Gamma_1\Gamma_0\right)|_{\zeta=\zeta_i}\Psi_i,\\
       \Psi_i[k]&=&\lim_{\delta\rightarrow0}\frac{[\delta+T_i[k]_{\zeta=\zeta_i}]\cdots[\delta+T_i[2]_{\zeta=\zeta_i}][\delta+T_i[1]|_{\zeta=\zeta_i}]\Gamma_{i-1}(\zeta_i+\delta)\cdots\Gamma_1(\zeta_i+\delta)\Gamma_0\Psi_i(\zeta_i+\delta)}{\delta^k}\\\\
       &=&\Psi_i+\sum_{s=1}^{k}\sum_{\substack{
m_i\geq h_1^{(i)}>\cdots>h_{k_i}^{(i)}\geq1,\\
i\geq g_1>\cdots>g_l\geq
1\\
\text{if $g_1=i$, then $h_1^{(1)}\leq
k$}}}^{\sum_{j=1}^{l}k_j+s=k}\left(T_{g_1}[h_1^{(1)}]\cdots
T_{g_1}[h_{k_1}^{(1)}]\cdots T_{g_l}[h_1^{(l)}]\cdots
T_{g_l}[h_{k_l}^{(l)}]\right)|_{\zeta=\zeta_i}\Psi_i^{[s]},\\
     \end{eqnarray*}
($1\leq k<m_i$),
then the following transformations
\begin{equation}
\label{nlsfields} \Psi[N]=T\Psi,\quad
q[N]=q+2\sum_{i=1}^{n}\sum_{j=1}^{m_i}(\zeta_i^*-\zeta_i)(P_i[m_j])_{21},
\quad (N=n+\sum_{k=1}^n m_k)
\end{equation}
constitute a generalized Darboux transformation for the NLS equation.
\end{thm}
We remark here that the solution formulae \eqref{n-fold} and \eqref{nlsfields}, represented in terms of summations, have certain merit. Indeed, for non-zero $\Psi_k, k=1, 2, \cdots, N$, all the denominators of $P[i]$ and $P_i[j]$ are easily seen to be  non-zero in these forms, therefore both  \eqref{n-fold} and \eqref{nlsfields} supply nonsingular solutions. The former could lead to $N$-soliton solutions, while the latter may yield rouge wave solutions.

Let us consider an example to illustrate the application of above formulae to construction of higher rogue wave solutions. To this
end, we  start with the seed
solution $q_0=e^{{\rm i}t}$. The corresponding solution for the linear spectral problem at $\zeta=ih$ is
\begin{equation}\label{nlssolution}
    \Psi_1(f)=\begin{pmatrix}
                   {\rm i}(C_1e^A-C_2e^{-A})e^{-\frac{{\rm i}}{2}t} \\
                   (C_2e^A-C_1e^{-A})e^{\frac{{\rm i}}{2}t} \\
                  \end{pmatrix}
\end{equation}
where
\[
C_1=\frac{(h-\sqrt{h^2-1})^{1/2}}{\sqrt{h^2-1}}, \quad C_2=\frac{(h+\sqrt{h^2-1})^{1/2}}{\sqrt{h^2-1}},\quad A=\sqrt{h^2-1}(x+{\rm i}ht).
\]
Let $h=1+f^2$, expanding the vector
function $ \Psi_1(f)$ at $f=0$, we have
\begin{equation}\label{seriesnls}
    \Psi_1(f)=\Psi_1(0)+\Psi_1^{[1]}f^2+\cdots
\end{equation}where
\begin{eqnarray*}
  \Psi_1(0) &=& \begin{pmatrix}
                  \left(-2t+2{\rm i}x-{\rm i}\right)e^{-\frac{1}{2}{\rm i}t} \\
                  \left(2{\rm i}t+2x+1\right)e^{\frac{1}{2}{\rm i}t} \\
                \end{pmatrix},
   \\
  \Psi_1^{[1]} &=&\begin{pmatrix}
                 \left[\frac{{\rm i}}{2}x-\frac{5}{2}t+\frac{{\rm i}}{4}-2tx^2+\frac{2{\rm i}}{3}x^3+\frac{2}{3}t^3-2{\rm i}xt^2-{\rm i}x^2+2tx+{\rm i}t^2\right] e^{-\frac{1}{2}{\rm i}t} \\
                 \left[\frac{1}{2}x+\frac{5{\rm i}}{2}t-\frac{1}{4}-2{\rm i}tx^2+\frac{2}{3}x^3-\frac{2{\rm i}}{3}t^3-2xt^2+x^2+2{\rm i}xt-t^2\right]e^{\frac{1}{2}{\rm i}t} \\
               \end{pmatrix}.
\end{eqnarray*}
It is clear that $\Psi_1(0)$ is a solution for
\eqref{nls_x}-\eqref{nls_t} at $\zeta={\rm i}$. By means of the formula \eqref{nlsn1}, we
obtain
\begin{align*}
    \Psi_1[1]&=\lim_{f\rightarrow0}\frac{[{\rm i}f^2+T_1[1]]\Psi_1(f)}{f^2}=T_1[1]\Psi_{1}^{[1]}+{\rm
    i}\Psi_1(0),\\
    T_1[1]&=2{\rm
    i}\left(I-\frac{\Psi_1(0)\Psi_1(0)^{\dag}}{\Psi_1(0)^{\dag}\Psi_1(0)}\right).
\end{align*}
Substituting above data into \eqref{nlsdt2} yields the second-order
rogue wave solution
\begin{equation*}\label{rogue2}
     q[2]=\left[1+\frac{G_1+{\rm i}tG_2}
    {H}\right]e^{{\rm
    i}t},\\
\end{equation*}
where
\begin{eqnarray*}
G_1&=&36-288x^2-192x^4-1152t^2x^2-864t^2-960t^4,\\
G_2&=&360+576x^2-192t^2-384x^4-768x^2t^2-384t^4,\\
H&=&64t^6+192t^4x^2+432t^4+396t^2+192t^2x^4-288t^2x^2+9+108x^2+64x^6+48x^4.
\end{eqnarray*}
which is nothing but the solution first constructed by Akhmediev {\it et al} \cite{ACA}. The higher order rouge wave solutions may be calculated similarly, thus we have a general approach to produce these solutions.

\subsection{Determinant forms and higher-order rogue wave}
In generic cases, iterated Darboux transformations may be given compactly by means of determinants and this is appealing mathematically. For the original Darboux transformation \eqref{nlsdt}, the result is well known \cite{Mat1}
\begin{thm}
Denoting $\Psi_i=(\psi_i,\phi_i)^T$ ($i=1, 2, \cdots N$),
then the $N$-fold Darboux transformation between fields \eqref{n-fold}
can be reformulated as
\begin{equation}\label{n-fold-fields1}
q[N]=q[0]-2\frac{\Delta_2}{\Delta_1},
\end{equation}
where
\begin{eqnarray*}
  \Delta_1 &=& \left|\begin{array}{cccccc}
                  \lambda_1^{N-1}\psi_1 &\cdots & \lambda_N^{N-1}\psi_N & -\lambda_1^{*(N-1)}\phi_1^* & \cdots& -\lambda_N^{*(N-1)}\phi_N^* \\
                  \cdots & \cdots & \cdots & \cdots & \cdots&\cdots  \\
                  \psi_1 & \cdots & \psi_N & -\phi_1^* &\cdots& -\phi_N^*   \\
                  \lambda_1^{N-1}\phi_1 & \cdots &  \lambda_N^{N-1}\phi_N & \lambda_1^{*(N-1)}\psi_1^*&\cdots & \lambda_{N}^{*(N-1)}\psi_N^*  \\
                  \cdots & \cdots & \cdots & \cdots & \cdots\cdots  \\
                  \phi_1 &  \cdots & \phi_N &\psi_1^* &\cdots& \psi_N^* \\
                \end{array}\right|, \\[8pt]
  \Delta_2 &=& \left|\begin{array}{cccccc}
                  \lambda_1^{N}\phi_1 &\cdots & \lambda_N^{N}\phi_N & \lambda_1^{*N}\psi_1^* &\cdots& \lambda_N^{*N}\psi_N^*  \\
                  \lambda_1^{N-2}\psi_1 &\cdots & \lambda_N^{N-2}\psi_N  & -\lambda_1^{*(N-2)}\phi_1^* & \cdots& -\lambda_N^{*(N-2)}\phi_N^*  \\
                  \cdots & \cdots & \cdots & \cdots & \cdots \cdots\\
                  \psi_1 &\cdots & \psi_N & -\phi_1^* &\cdots&  -\phi_N^* \\
                  \lambda_1^{N-1}\phi_1 &\cdots & \lambda_N^{N-1}\phi_N &  \lambda_1^{*(N-1)}\psi_1^* &\cdots& \lambda_N^{*(N-1)}\psi_N^* \\
                  \cdots & \cdots & \cdots & \cdots & \cdots\cdots \\
                  \phi_1 & \cdots & \phi_N & \psi_1^* & \cdots&\psi_N^* \\
                \end{array}\right|.
\end{eqnarray*}
\end{thm}

To find the determinant representations of our generalized Darboux transformation, we observe that the limit process
presented in above subsection may be taken directly, namely
\eqref{nlsdt2} may be obtained by the following consideration
\begin{eqnarray*}
   q[2] &=& q[1]+\lim_{\zeta_2\rightarrow \zeta_1}
  2(\zeta_2^*-\zeta_2)(P[2])_{21}.
\end{eqnarray*}
Thus, as for the KdV case worked out in \cite{Mat1}, we may perform the limit on the determinant form  \eqref{n-fold-fields1} and get
 \begin{thm}
 Assuming that  $N$ distinct
solutions $\Psi_i=(\psi_i,\phi_i)^T$ $(i=1,2,\cdots,n)$ are given for the spectral problem \eqref{nls_x}\eqref{nls_t} at   $\zeta=\zeta_1, \cdots, \zeta=\zeta_n$ and expanding
 \begin{eqnarray*}
  (\zeta_i+\delta)^j\psi_i(\zeta_i+\delta) &=&\zeta_i^j\psi_i+\psi_i[j,1]\delta+\psi_i[j,2]\delta^2+\cdots+\psi_i[j,m_i]\delta^{m_i}+\cdots,\\
  (\zeta_i+\delta)^j\phi_i(\zeta_i+\delta)
  &=&\zeta_i^j\phi_i+\phi_i[j,1]\delta+\phi_i[j,2]\delta^2+\cdots+\phi_i[j,m_i]\delta^{m_i}+\cdots,
\end{eqnarray*}
with
\[
  \psi_i[j,m]=\frac{1}{m!}\frac{\partial^m}{\partial
\zeta^m}[\zeta^j\psi_i(\zeta)]|_{\zeta=\zeta_i},  \quad
\phi_i[j,m]=\frac{1}{m!}\frac{\partial^m}{\partial
\zeta^m}[\zeta^j\phi_i(\zeta)]|_{\zeta=\zeta_i},
\]
($j=0,1,\cdots,n$, $m=1,2,3,\cdots$), then we have
\begin{equation}\label{nlsdeter}
    q[N]=q-2\frac{D_2}{D_1},\quad D_2=\det([H_1\cdots H_n]),\quad \quad D_1=\det([G_1\cdots
    G_n]),
\end{equation}
where $N=n+\sum_{k=1}^n m_k$ and
\begin{eqnarray*}
        G_i &=& \begin{bmatrix}
                  \zeta_i^{N-1}\psi_i &\cdots & \psi_i[N-1,m_i] & -\zeta_i^{*(N-1)}\phi_i^* & \cdots& -\phi_i[N-1,m_i]^{*} \\
                  \cdots & \cdots & \cdots & \cdots & \cdots&\cdots  \\
                  \psi_i & \cdots & \psi_i[0,m_i] & -\phi_i^* &\cdots& -\phi_i[0,m_i]^*  \\
                  \zeta_i^{N-1}\phi_i & \cdots & \phi_i[N-1,m_i] & \zeta_i^{*(N-1)}\psi_i^*&\cdots & \psi_i[N-1,m_i]^*  \\
                  \cdots & \cdots & \cdots & \cdots & \cdots\cdots  \\
                  \phi_i &  \cdots & \phi_i[0,m_i] &\psi_i^* &\cdots& \psi_i[0,m_i]^* \\
                \end{bmatrix},
         \\
        H_i &=& \begin{bmatrix}
                  \zeta_i^{N}\phi_i &\cdots & \phi_i[N,m_i] & \zeta_i^{*N}\psi_i^* &\cdots&  \psi_i[N,m_i]^{*}  \\
                  \zeta_i^{N-2}\psi_i &\cdots & \psi_i[N-2,m_i] & -\zeta_i^{*(N-2)}\phi_i^* & \cdots& -\phi_i[N-2,m_i]^{*}  \\
                  \cdots & \cdots & \cdots & \cdots & \cdots \cdots\\
                  \psi_i &\cdots & \psi_i[0,m_i] & -\phi_i^* &\cdots&  -\phi_i[0,m_i]^* \\
                  \zeta_i^{N-1}\phi_i &\cdots & \phi_i[N-1,m_i] &  \zeta_i^{*(N-1)}\psi_i^* &\cdots& \psi_i[N-1,m_i]^* \\
                  \cdots & \cdots & \cdots & \cdots & \cdots\cdots \\
                  \phi_i & \cdots & \phi_i[0,m_i] & \psi_i^* & \cdots&\psi_i[0,m_i]^* \\
                \end{bmatrix}.
      \end{eqnarray*}
\end{thm}

We point out that, applying to special seed solutions, \eqref{nlsdeter} enables us to have a determinant form for higher order rouge wave solutions. We consider
\begin{equation}\label{solution3}
     \psi_1 = {\rm i}(C_1e^A-C_2e^{-A}), \quad
   \phi_1 = (C_2e^A-C_1e^{-A}),
\end{equation}
where\[
\begin{array}{ll}
C_1=\frac{(1+f^2-f\sqrt{2+f^2})^{1/2}}{f\sqrt{2+f^2}},& C_2=\frac{(1+f^2+f\sqrt{2+f^2})^{1/2}}{f\sqrt{2+f^2}},\\
A=f\sqrt{2+f^2}[x+{\rm i}(1+f^2)t+\Phi(f)],&
\Phi(f)=\sum_{i=0}^{N}s_if^{2i},\quad s_i\in \mathds{C}.
\end{array}
\]
The associated Taylor expansions are
\begin{eqnarray*}
     {\rm i}^j(1+f^2)^{j}\psi_1(f) &=& {\rm
     i}^{j}\psi_1(0)+\psi_1[j,1]f^2+\cdots+\psi_1[j,N]f^{2N}+\cdots,\\ \psi_1[j,n]&=&\frac{1}{(2n)!}\frac{\partial^{2n}}{\partial
f^{2n}}[{\rm i}^j(1+f^2)^{j}\psi_1(f)]|_{f=0}, \\
     {\rm i}^j(1+f^2)^{j}\phi_1(f) &=& {\rm
     i}^{j}\phi_1(0)+\phi_1[j,1]f^2+\cdots+\phi_1[j,N]f^{2N}+\cdots,\\ \phi_1[j,n]&=&\frac{1}{(2n)!}\frac{\partial^{2n}}{\partial
f^{2n}}[{\rm i}^j(1+f^2)^{j}\phi_1(f)]|_{f=0},
\end{eqnarray*}
($j=0,1,\cdots,N,$ $n=1,2,3,\cdots$). It follows that the $N$-th order
rogue wave solution for NLS equation \eqref{nls} reads as
\begin{equation}\label{grogue}
     q[N] = \left[1-2\frac{D_2}{D_1}\right]e^{{\rm i}t},
\end{equation}
where
\begin{eqnarray*}
   D_1 &=& \begin{vmatrix}
                  {\rm i}^{N-1}\psi_1 &\cdots & \psi_1[N-1,N-1] & -{\rm (-i)}^{(N-1)}\phi_1^* & \cdots& -\phi_1[N-1,N-1]^{*} \\
                  \cdots & \cdots & \cdots & \cdots & \cdots&\cdots  \\
                  \psi_1 & \cdots & \psi_1[0,N-1] & -\phi_1^* &\cdots& -\phi_1[0,N-1]^*  \\
                  {\rm i}^{N-1}\phi_1 & \cdots & \phi_1[N-1,N-1] & {\rm -i}^{(N-1)}\psi_1^*&\cdots & \psi_1[N-1,N-1]^*  \\
                  \cdots & \cdots & \cdots & \cdots & \cdots\cdots  \\
                  \phi_1 &  \cdots & \phi_1[0,N-1] &\psi_1^* &\cdots& \psi_1[0,N-1]^* \\
                \end{vmatrix},
         \\[8pt]
        D_2 &=& \begin{vmatrix}
                  {\rm i}^{N}\phi_1 &\cdots & \phi_1[N,N-1] & {\rm (-i)}^{N}\psi_1^* &\cdots&  \psi_1[N,N-1]^{*}  \\
                  {\rm i}^{N-2}\psi_1 &\cdots & \psi_1[N-2,N-1] & -{\rm (-i)}^{(N-2)}\phi_1^* & \cdots& -\phi_1[N-2,N-1]^{*}  \\
                  \cdots & \cdots & \cdots & \cdots & \cdots \cdots\\
                  \psi_1 &\cdots & \psi_1[0,N-1] & -\phi_1^* &\cdots&  -\phi_1[0,N-1]^* \\
                  {\rm i}^{N-1}\phi_1 &\cdots & \phi_1[N-1,N-1] &  {\rm (-i)}^{(N-1)}\psi_1^* &\cdots& \psi_1[N-1,N-1]^* \\
                  \cdots & \cdots & \cdots & \cdots & \cdots\cdots \\
                  \phi_1 & \cdots & \phi_1[0,N-1] & \psi_1^* & \cdots&\psi_1[0,N-1]^* \\
                \end{vmatrix}.
\end{eqnarray*}

For the case $N=2$, above formula may provide the 2nd order rogue wave solution with two free parameters for the NLS equation, which has been analyzed in detail in \cite{ADN}. It is shown that this solution splits into three 1st order rogue waves rather than two.  Indeed, our results supply the high order rogue
wave solutions with more free parameters, which determine the
spatial-temporal structures of the solutions. In particular, the 3rd rogue wave solution, possessing four free
parameters, may be worked out by setting  $N=3$ and $\Phi(f)=(b+{\rm i}c)f^2+(e+{\rm i}g)f^4$ in our formula, whose explicit expression is omitted since it is rather cumbersome. In the following, we consider two special cases, which have different
spatial-temporal patterns.

\begin{itemize}
\item {Case A}

In this case, we assume $b=c=0$, $|e| \gg 0$ or $|g| \gg
0$. The corresponding 3rd order rogue wave solution is composed of six 1st order rogue
waves, which array a
regular pentagon. Interestingly, among the six 1st order rogue
waves, one sits on the center and the rest locate on the vertices of the pentagon.
After some calculations, we find that the radial distance from the center of the pentagon approximately equals to  $\frac{1}{2}360^{1/5}
(e^2+g^2)^{1/10}$. For the case $g=0$ and $e\gg
0$, one of the vertices locates on the negative direction of $x$ axis and the corresponding quadrantal angle for the $(e,g)$-plane is assumed to be zero. For the general non-zero $g$, the pentagon
 will rotate $\frac{1}{5}\theta$ along anticlockwise, where
$\theta$ is the associated quadrantal angle for the $(e,g)$-plane.
\begin{figure}
        \subfigure[The third-order rogue wave solution $|q|^2$]{
        \begin{minipage}[b]{0.4\textwidth}
          \centering
       \includegraphics[width=3in]{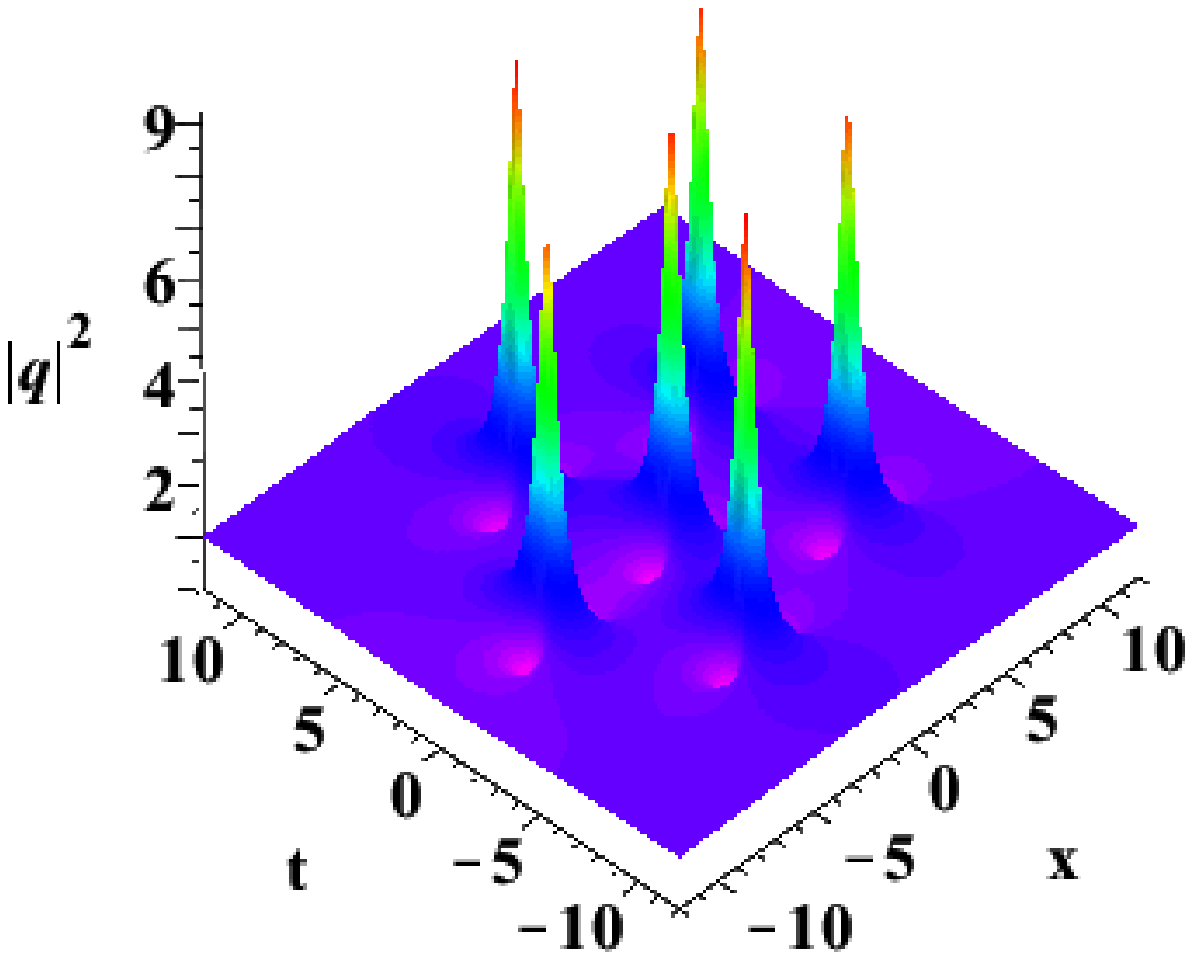}
        \end{minipage}}%
        \hspace{0.04\textwidth}%
       \subfigure[Density plot for the third-order rogue wave solution
$|q|^2$ ]{ \begin{minipage}[b]{0.4\textwidth}
          \centering
       \includegraphics[width=2.2in]{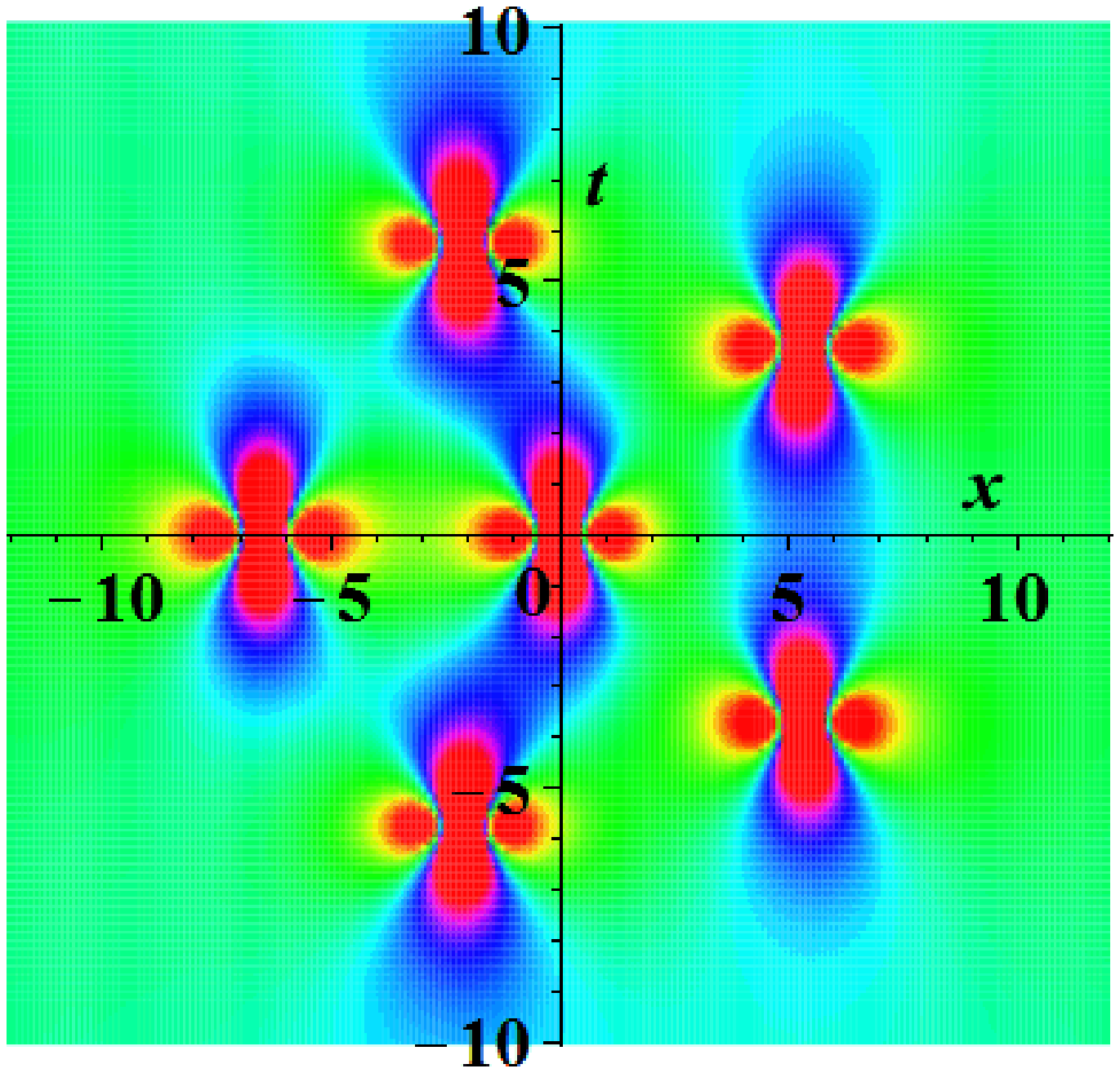}
        \end{minipage}}
\caption{The parameters $e=1000$ and $g=0$}\label{p2}
      \end{figure}

\item Case B

 For the second case, we take the parameters
$e=g=0$, $|b| \gg 0$ or $|c|\gg 0$. The corresponding third order rogue wave
consists of the six first order rogue waves as well, which array an equilateral triangle.
 The distance between the center and any vertex approximately equals to $90^{1/6}(b^2+c^2)^{1/6}$ . For the case $c=0$ and $b\gg 0$, one of the
vertices locates on the positive direction of $x$ axis and and the corresponding quadrantal angle for the $(b,c)$-plane is assumed to be zero. For the general non-zero $c$, the triangle will
rotate $\frac{1}{3}\theta$ anticlockwisely, where $\theta$ is
the related quadrantal angle for the $(b,c)$-plane.
When
$b=c=e=g=0$, the rogue wave is the one considered by Akhmediev {\em et al } in \cite{AAS}.
\begin{figure}
        \subfigure[The third-order rogue wave solution $|q|^2$]{
        \begin{minipage}[b]{0.4\textwidth}
          \centering
       \includegraphics[width=3in]{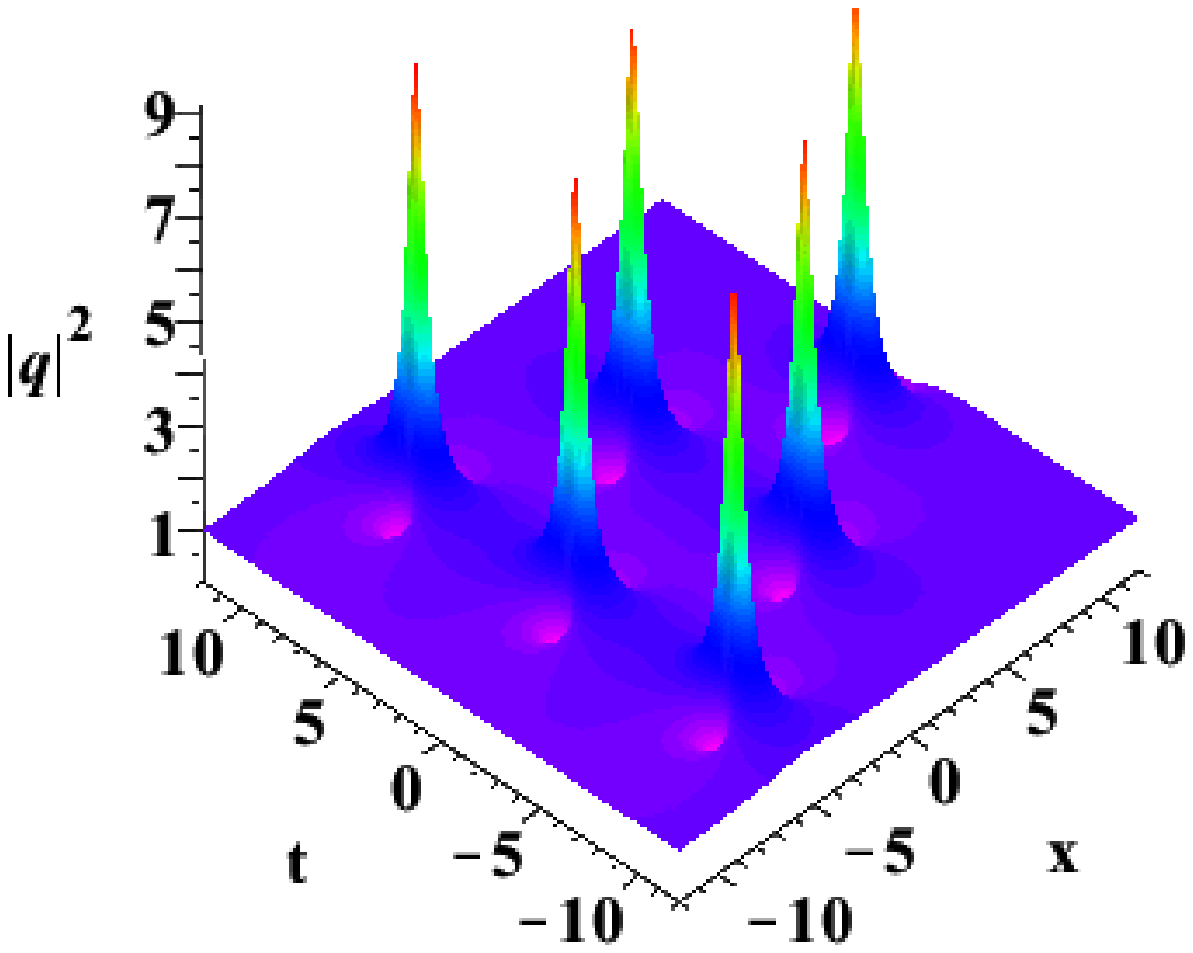}
        \end{minipage}}%
        \hspace{0.04\textwidth}%
       \subfigure[Density plot for the third-order rogue wave solution
$|q|^2$ ]{ \begin{minipage}[b]{0.4\textwidth}
          \centering
       \includegraphics[width=2.2in]{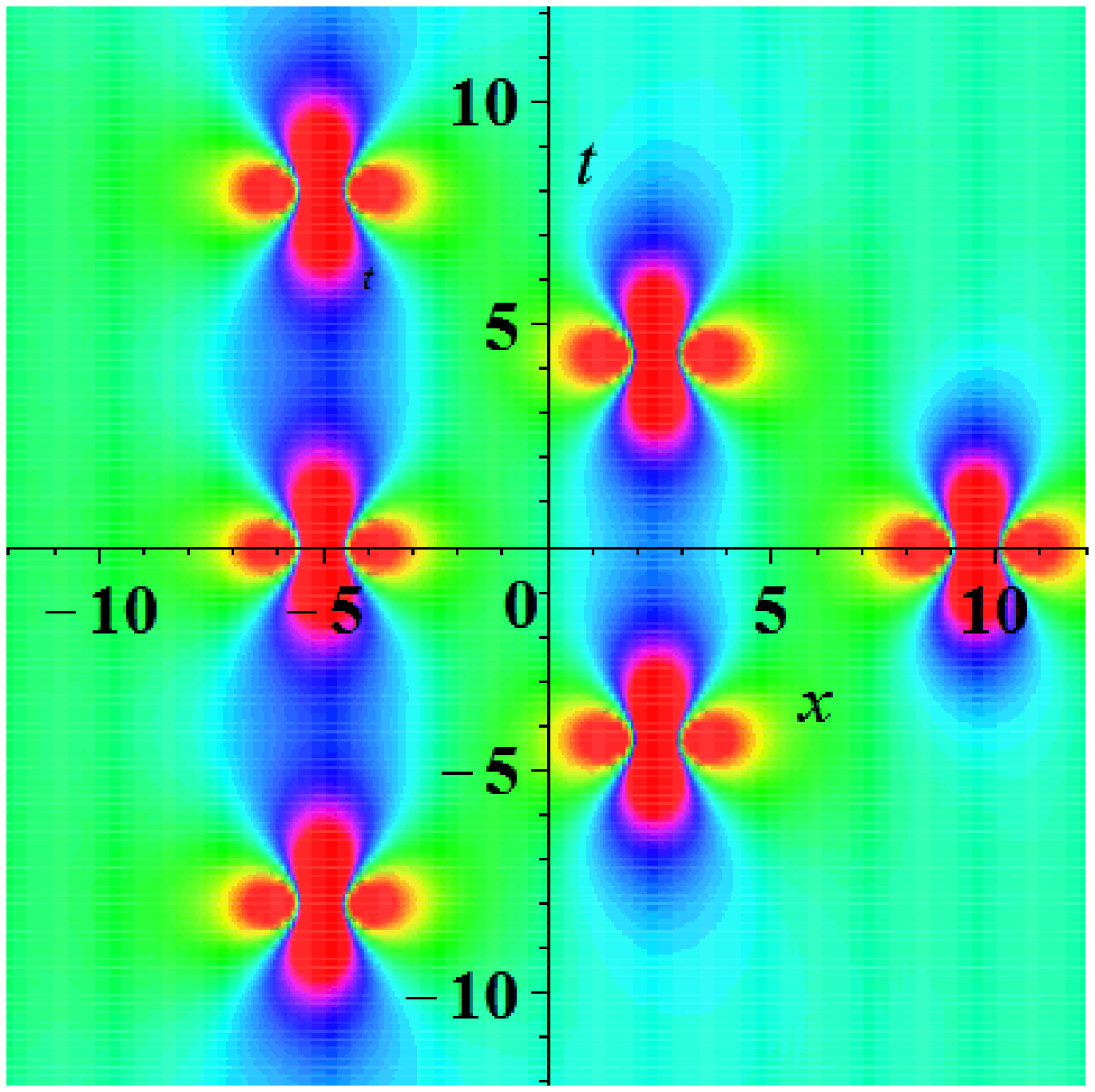}
        \end{minipage}}
\caption{The parameters $b=100$ and $c=0$}\label{p1}
      \end{figure}
\end{itemize}

\begin{figure}
        \subfigure[The fourth-order rogue wave solution $|q|^2$]{
        \begin{minipage}[b]{0.4\textwidth}
          \centering
       \includegraphics[width=3in]{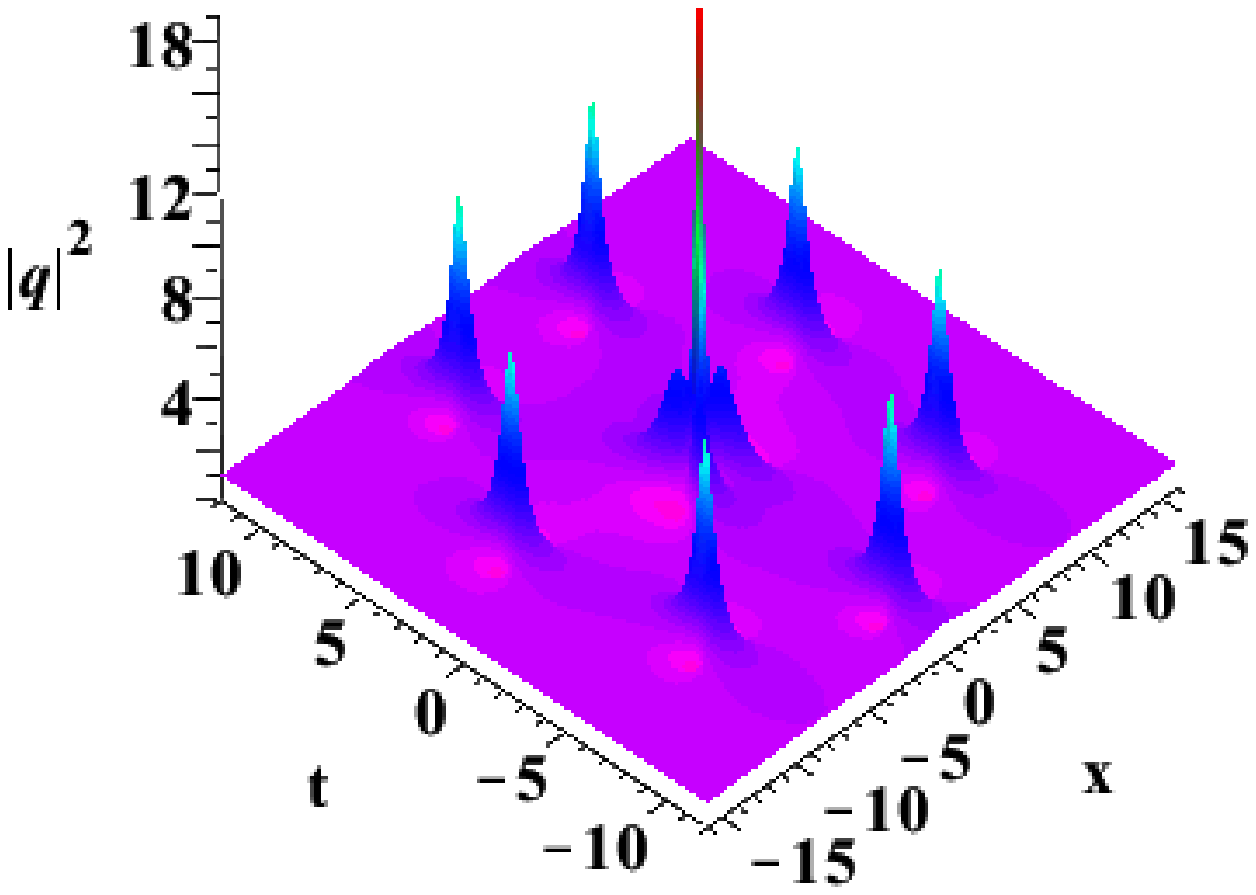}
        \end{minipage}}%
        \hspace{0.04\textwidth}%
       \subfigure[Density plot for the fourth-order rogue wave solution
$|q|^2$ ]{ \begin{minipage}[b]{0.4\textwidth}
          \centering
       \includegraphics[width=2.2in]{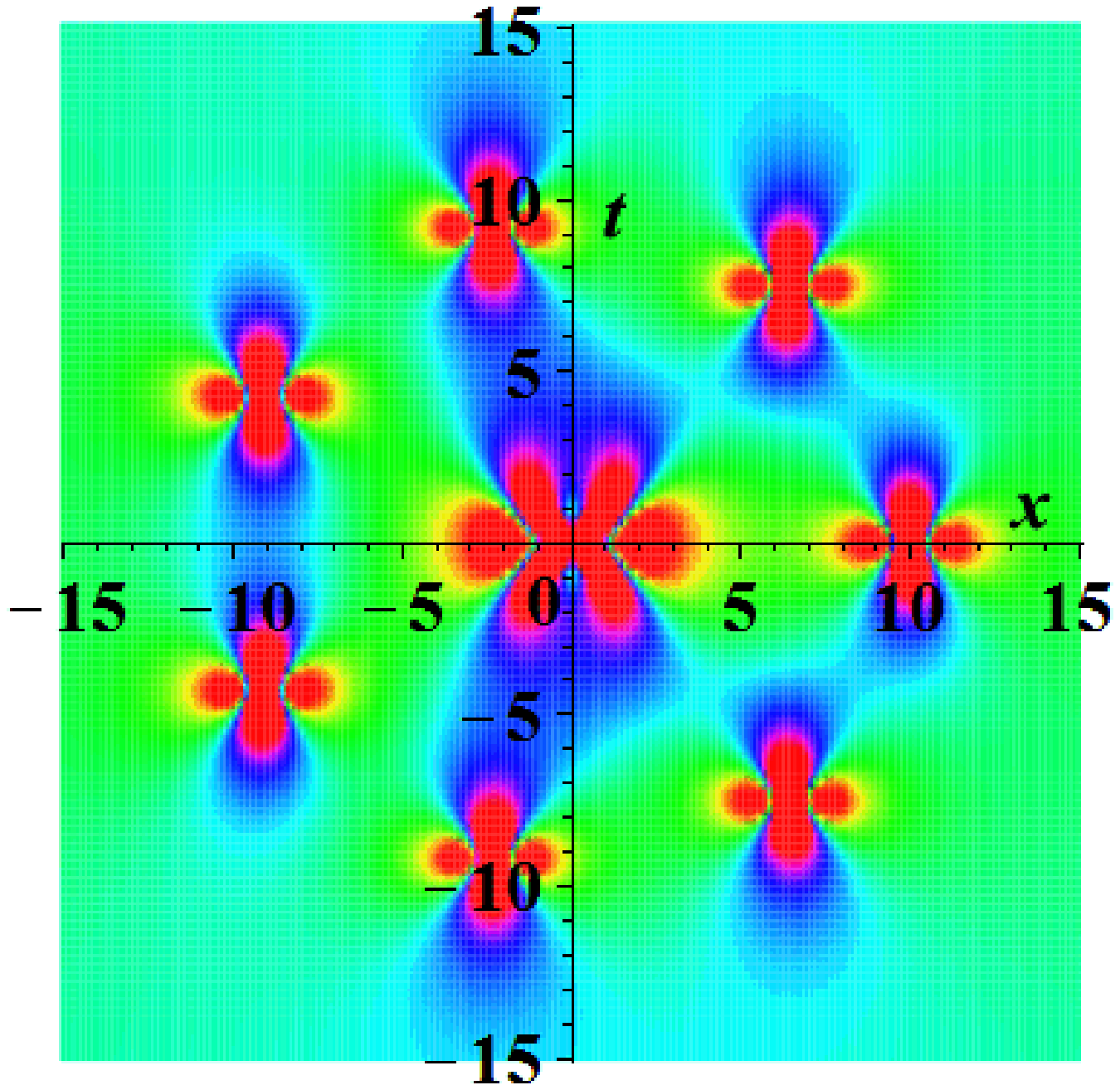}
        \end{minipage}}
\caption{The parameters $\Phi(f)=5\times 10^5{\rm
i}f^{6}$}\label{p3}
      \end{figure}

Let $(x_i,t_i)$ $(i=1,...,6)$ be the coordinates of the six peaks, then we find that the third order rogue wave may be approximated by
\begin{equation*}
    q[3]\approx\sum_{i=1}^6\left[-1+4\frac{1+2{\rm i}(t-t_i)}{d_i}\right]e^{{\rm
    i}t},\quad d_i=1+4(x-x_i)^2+4(t-t_i)^2,
\end{equation*}
 the
``center of mass'' is at the origin $(0,0)$ in both cases. With raising of the order, the rogue solution  contains more free parameters and exhibits  more
interesting spatial-temporal structures. For instance, choosing the proper parameters, the 4th
order rogue wave possesses the regular heptagon spatial-temporal
pattern (Fig. \ref{p3}), and a second order
rogue wave locates on the center. Naturally, we  conjecture that the spatial-temporal
pattern of N-th order rogue wave possesses $2N-1$-gon
spatial-temporal pattern by choosing the parameters
$\Phi(f)=cf^{2(N-1)}$ ($|c|\gg0$) in formula \eqref{grogue}, where $c$ is a
complex number.

\begin{figure}
        \subfigure[The third-order rogue wave solution $|q|^2$]{
        \begin{minipage}[b]{0.4\textwidth}
          \centering
       \includegraphics[width=3in]{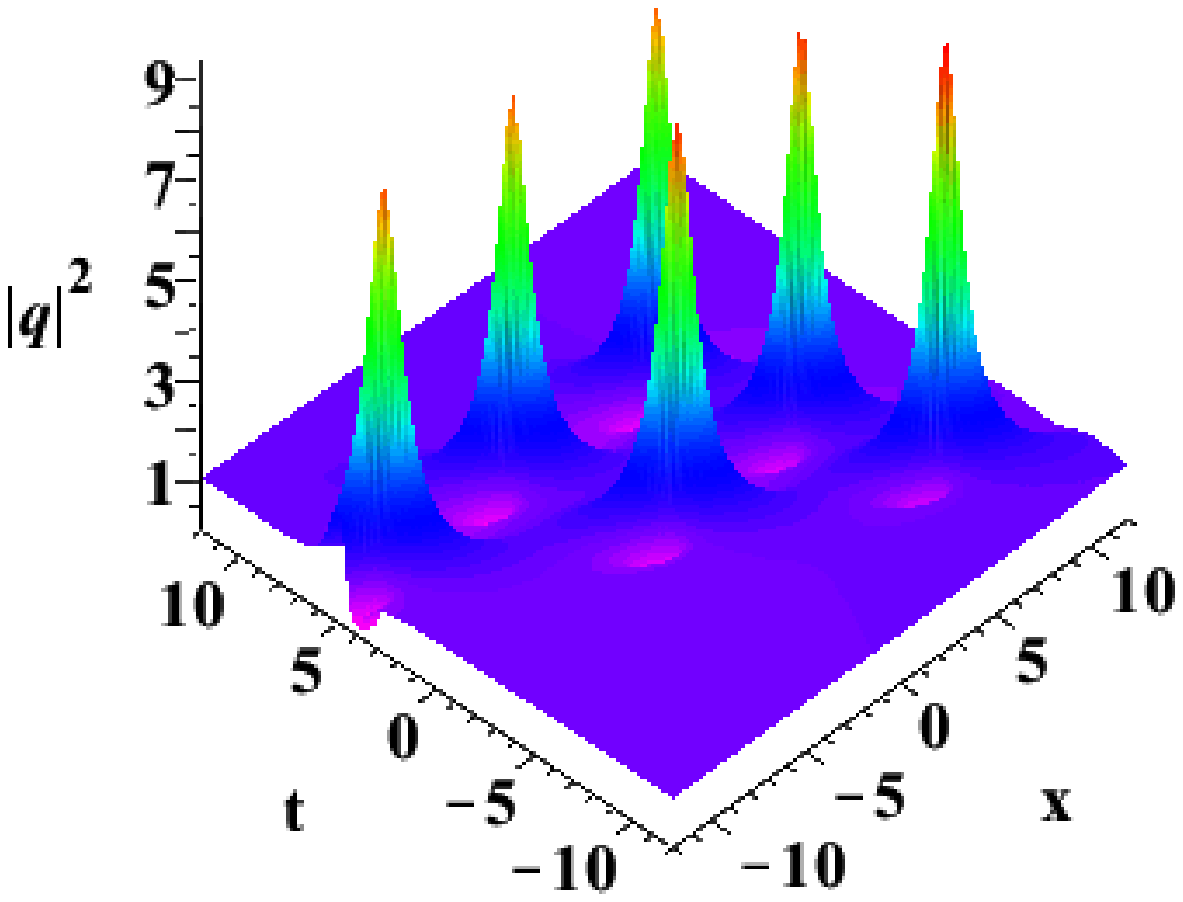}
        \end{minipage}}%
        \hspace{0.04\textwidth}%
       \subfigure[Density plot for third-order rogue wave solution $|q|^2$]{ \begin{minipage}[b]{0.4\textwidth}
          \centering
       \includegraphics[width=2.2in]{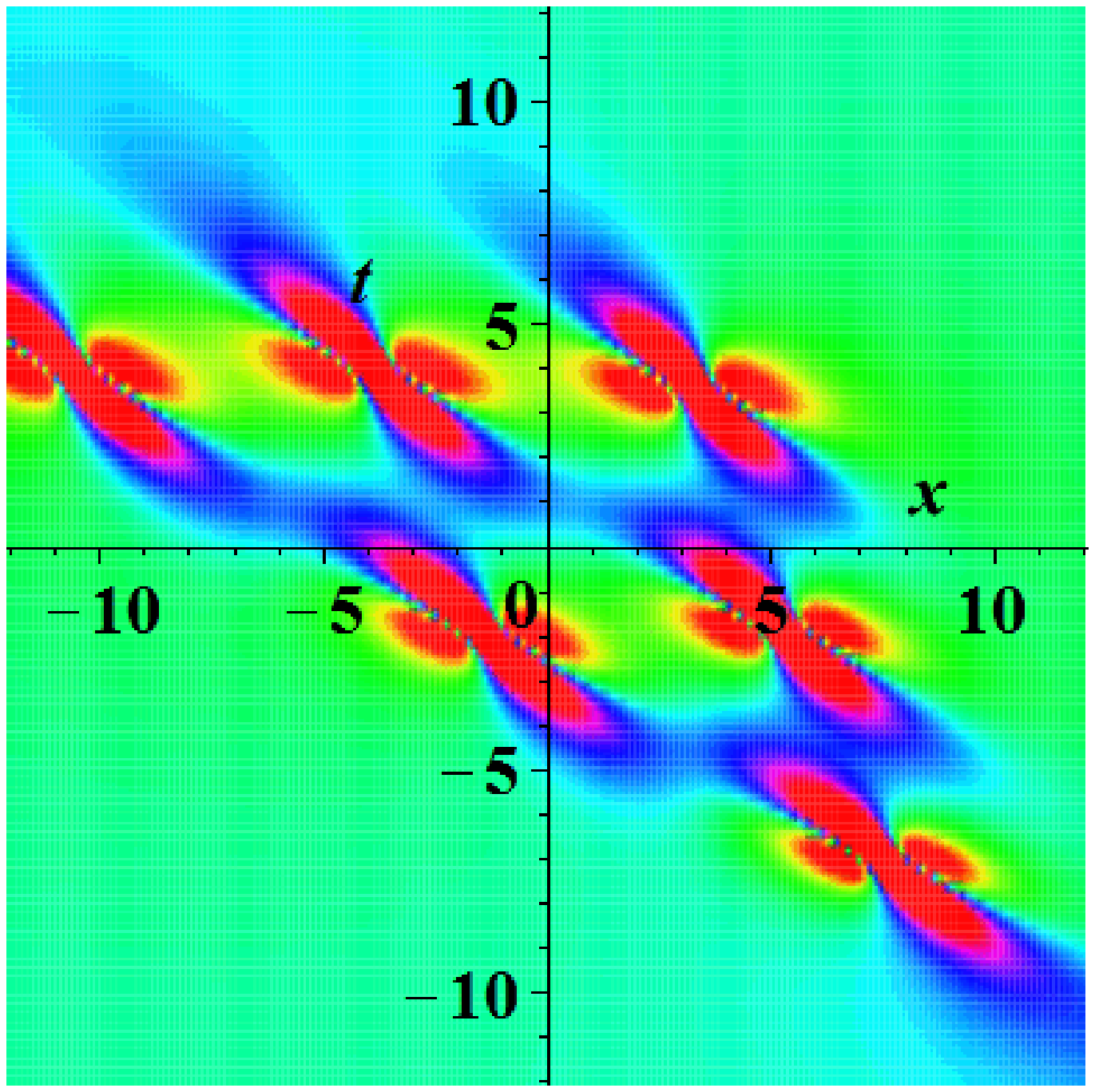}
        \end{minipage}}
\caption{The parameters $\alpha=\frac{1}{6}$, $\Phi(f)=50{\rm
i}f^2$.}\label{p4}
      \end{figure}

We conclude this section with the following remarks in order:
\begin{enumerate}
\item If we choose the seed solution $q[0]=0$, we may obtain the higher
soliton solution \cite{GS,SY}.
\item
The integrable Hirota equation \begin{equation}\label{He}
     {\rm i}q_t+\frac{1}{2}q_{xx}+|q|^2q-{\rm
     i}\alpha(q_{xxx}+6|q|^2q_x)=0,\quad \text{$\alpha$ is real constant}
\end{equation} is the third flow of NLS
hierarchy. Its rogue wave solutions and rational solutions were
discussed in \cite{ASA}. To obtain the $N$-th order rogue wave
solution for Hirota equation \eqref{He} from \eqref{grogue}, we
merely need to modify the $A=f\sqrt{2+f^2}[x+{\rm
i}(1+f^2)t+\alpha(2+4(1+f^2)^2)t+\Phi(f)]$ in \eqref{solution3}.
Rather than giving the explicit exprssions, which is lengthy, we
plot the third-order rogue wave solution for Hirota equation with
the Fig. \ref{p4}. Due to the third-order dispersion and time-delay
correction to the cubic term \cite{ASA}, the solution enjoys the
different behavior from NLS equation.
\end{enumerate}

\section{Conclusion}
Through a limit procedure, we have generalized the original Darboux
transformation for the NLS equation. This Darboux transformation, in
particular, allows us  to calculate higher order rogue wave
solutions in a unified way. We believe that the idea is rather
general and could be applied to other physically interested models
as well. These spatial-temporal structure for N-th order rogue wave
may be useful to research spatial-temporal distributing of rogue
wave in deep water.

\textbf{Acknowledgment}. This work is supported by the National
Natural Science Foundation of China (grant numbers:
10731080, 10971222) and the Fundamental Research Funds for Central
Universities.

\end{document}